\documentstyle[aps,preprint,tighten,epsfig,amssymb]{revtex}

\begin{document}

\draft

\preprint{
\vbox{\hbox{SNUTP-00-001} \hbox{FTUV-00-0118} \hbox{hep-ph/0001209} }}

\title{Anomalous $f_1$ exchange in vector meson\\
       photoproduction asymmetries}

\author{
Yongseok Oh$^a$%
\footnote{e-mail address: yoh@mulli.snu.ac.kr},
Nikolai I. Kochelev$^{a,b}$%
\footnote{e-mail address: kochelev@thsun1.jinr.ru},
Dong-Pil Min$^{a}$%
\footnote{e-mail address: dpmin@mulli.snu.ac.kr},
Vicente Vento$^c$%
\footnote{e-mail address: vicente.vento@uv.es},
\\ and
Andrey V. Vinnikov$^{d}$%
\footnote{e-mail address: vinnikov@thsun1.jinr.ru}}

\address{
\bigskip
$^a$ Department of Physics and Center for Theoretical Physics,
     Seoul National University, Seoul 151-742, Korea\\
$^b$ Bogoliubov Laboratory of Theoretical Physics,
     JINR, Dubna, Moscow region, 141980 Russia\\
$^c$ Departament de F\'{\i}sica Te\`orica and Institut de F\'{\i}sica
     Corpuscular, Universitat de Val\`encia-CSIC E-46100
     Burjassot (Valencia), Spain \\
$^d$ Far Eastern State University, Sukhanova 8, GSP, Vladivostok,
     690660 Russia}

\date{\today}

\maketitle

\begin{abstract}

We perform an analysis of the elastic production of vector mesons 
with polarized photon beams at high energy in order to investigate
the validity of a recently proposed dynamical mechanism based on the
dominance of the $f_1$ trajectory at large momentum transfer.
The density matrix characterizing the angular distributions of the
vector meson decays is calculated within an exchange model which
includes the Pomeron and the $f_1$. 
The asymmetries of these decays turn out to be very useful to
disentangle the role of these exchanges since their effect depends
crucially on their quantum numbers which are different.
The observables analyzed are accessible with present experimental
facilities.

\end{abstract}


Exclusive photoproduction experiments of vector mesons have become
powerful tools for testing diffractive mechanisms at high energy 
\cite{BSYP78,Crit97,AC99}.
Regge theory has been successful in describing the diffractive
production in terms of the Pomeron exchange mechanism.
Donnachie and Landshoff \cite{DL87a,DL88a} showed that, by
assuming the Pomeron-photon analogy and introducing a form factor 
for the coupling of the Pomeron to quarks, the diffractive vector 
meson production with real and virtual photons could be described
successfully by the soft Pomeron exchange.
The soft Pomeron exchange governs this process for small $|t|$ 
and fulfills the $s$-channel helicity conservation, a consequence 
of the old vector dominance model \cite{Saku69} and a requirement of
the experimental data \cite{BCGG72,BCEK73}. 
However, at larger $|t|$, the soft Pomeron alone cannot explain
the recent ZEUS data on elastic vector meson photo- and 
electroproduction \cite{ZEUS99} and new contributions seem 
necessary.
For example Donnachie and Landshoff \cite{DL98,DL99} introduce in
addition to the {\em soft\/} Pomeron the {\em hard\/} Pomeron with
the trajectory $\alpha_{P'}^{} = 1.44 + 0.1 t$ describing in this way
the data up to $|t| \sim 2$ GeV$^2$.

Recently we have suggested a new anomalous Regge trajectory with 
high intercept $\alpha_{f_1}^{}(0) \approx 1$ and small slope
$\alpha^\prime_{f_1} \approx 0$ \cite{KMOV99}.%
\footnote{It is not well known how the Pomeron arises from QCD, although 
it seems quite plausible that it is related to the conformal anomaly 
of the theory \cite{KL99,Shu00}.
In the same way we do not yet know how the anomalous $f_1$ trajectory
arises from QCD, but we have strong suspicions that the origin of its
physical relevance lies in its relation to the {\it axial anomaly} of
the theory.}
This trajectory has the quantum numbers $P = C = +1$ and the signature
$\sigma = -1$ while the Pomeron carries $P = C = \sigma = +1$.
In Ref. \cite{KMOV99} we have shown that the $f_1$ exchange describes
the vector meson photoproduction data at large energy and momentum
transfer.%
\footnote{The $f_1$ trajectory also gives natural explanation to the
behavior of the cross sections of elastic hadron-hadron scattering at
large $|t|$ and furthermore its contribution to the flavor singlet part
of the spin-dependent structure function $g_1$ at low $x$ region gives
a new explanation to the proton spin problem.}
In this model, the soft Pomeron is dominant at $|t| \le 1$ GeV$^2$
while the $f_1$ exchange dominates the large $|t|$ region,
$|t| \ge 1$ GeV$^2$.

In order to understand the details of the mechanisms involved in our
model it is important to investigate other physical quantities which
can distinguish between the two exchanges, i.e., the Pomeron and the
$f_1$, in a clear way.
Diffractive production of vector mesons by polarized photon beams seems
to be the appropriate tool for such purpose as we will show hereafter.

One of the important features of the new anomalous $f_1$ trajectory is
its odd signature, which should discriminate it from the Pomeron which
has even signature.
Therefore the contribution from this new exchange can be disentangled
from the Pomeron contribution in spin-dependent processes.
In order to investigate the very specific features of the $f_1$
trajectory contribution we consider vector meson production with
polarized photon beams and its decay into pseudoscalar mesons.
We find that the asymmetries of the vector meson decays described by
the soft Pomeron and $f_1$ exchanges are drastically different from the
predictions obtained with the soft and hard Pomeron exchanges.

Our starting point is the density matrix of the vector meson 
production by photons from proton targets,
\begin{equation}
\rho_{\lambda_V^{} \lambda_V'} = \frac{1}{N}
\sum_{\lambda_N',\lambda_\gamma,\lambda_N^{},\lambda_\gamma'}
T_{\lambda_V^{} \lambda_N', \lambda_\gamma \lambda_N^{}}
\rho(\gamma)_{\lambda_\gamma \lambda_\gamma'}
T^*_{\lambda_V' \lambda_N', \lambda_\gamma' \lambda_N^{}},
\end{equation}
where $T$ is the $T$-matrix element of elastic vector meson
photoproduction process, $\lambda$'s are the polarization states of
the particles, and $N$ is the normalization factor defined as
\begin{equation}
N = \frac12 \sum_{\mbox{\scriptsize $\lambda$'s}}
|T_{\lambda_V^{} \lambda_N', \lambda_\gamma \lambda_N^{}}|^2.
\end{equation}
The photon density matrix $\rho(\gamma)$ is given by
\begin{equation}
\rho(\gamma) = \frac12 ( 1 + {\bf P}_\gamma \cdot \bbox{\sigma} ),
\label{DMphoton}
\end{equation}
where $\bbox{\sigma}$ is the Pauli matrix and ${\bf P}_\gamma$
specifies the polarization of linearly polarized photons and is given by
\begin{equation}
{\bf P}_\gamma = p_\gamma (-\cos 2\Phi, -\sin 2\Phi, 0),
\end{equation}
where $\Phi$ denotes the angle between the photon polarization vector
and the vector meson production plane, and $p_\gamma$ denotes the
magnitude of the polarization ($0 \le p_\gamma \le 1$).
The decay angular distribution of the vector meson in its rest
frame reads
\begin{eqnarray}
\frac{d \mathcal{N}}{d\cos\vartheta d\varphi} \equiv
W(\cos\vartheta,\varphi,\Phi) =
W^0 (\cos\vartheta, \varphi) + \sum_{i=1}^3
P^i_\gamma (\Phi) W^i (\cos\vartheta, \varphi),
\label{decay}
\end{eqnarray}
where $\vartheta$ and $\varphi$ are the polar and azimuthal angles of
the direction of flight of one pseudoscalar meson in the vector meson
rest frame.
As in the literature, we use the Gottfried-Jackson frame \cite{GJ64} as
the vector meson rest frame, where the $z$ axis is in the direction of
the incident photon as seen in this frame.
(See Refs. \cite{BSYP78,SSW70,SW73,WS78} for details.)

The explicit forms of $W^\alpha$ are
\begin{eqnarray}
W^0 (\cos\vartheta, \varphi) &=& \frac{3}{4\pi} \Bigl\{
\frac12 ( 1 - \rho^0_{00} ) + \frac12 ( 3 \rho^0_{00} - 1 )
\cos^2\vartheta
- \sqrt2 \, \mbox{Re} \rho^0_{10} \sin2\vartheta \cos\varphi
\nonumber \\ && \qquad \mbox{}
- \mbox{Re} \rho^0_{1-1} \sin^2\vartheta \cos2\varphi \Bigr\},
\nonumber \\
W^1 (\cos\vartheta, \varphi) &=& \frac{3}{4\pi} \Bigl\{
\rho^1_{11} \sin^2\vartheta + \rho^1_{00} \cos^2\vartheta
- \sqrt2 \, \mbox{Re} \rho^1_{10} \sin2\vartheta \cos\varphi
- \mbox{Re} \rho^1_{1-1} \sin^2\vartheta \cos2\varphi \Bigr\},
\nonumber \\
W^2 (\cos\vartheta, \varphi) &=& \frac{3}{4\pi} \Bigl\{
\sqrt2 \, \mbox{Im} \rho^2_{10} \sin2\vartheta \sin\varphi
+ \mbox{Im} \rho^2_{1-1} \sin^2\vartheta \sin2\varphi \Bigr\},
\nonumber \\
W^3 (\cos\vartheta, \varphi) &=& \frac{3}{4\pi} \Bigl\{
\sqrt2 \, \mbox{Im} \rho^3_{10} \sin2\vartheta \sin\varphi
+ \mbox{Im} \rho^3_{1-1} \sin^2\vartheta \sin2\varphi \Bigr\},
\end{eqnarray}
where $\rho^\alpha_{ij}$ are the matrix elements of $\rho^\alpha$ which
are defined by
\begin{equation}
\rho^0 = \frac{1}{2N} TT^\dagger, \quad
\rho^i = \frac{1}{2N} T \sigma^i T^\dagger,
\label{dmdef}
\end{equation}
with $i = 1,2,3$, whose normalization is $\mbox{Tr} \rho^0 = 1$.
There are various decay angular distribution functions arising from
different photon polarizations, whose measurements determine the vector
meson density matrix elements.
Interesting quantities in connection with the nature of the exchanged
particles are the asymmetries.

Depending on the direction of the polarization vector of the linearly
polarized photon beams, we define the asymmetry $\Sigma$ as
\begin{equation}
\Sigma \equiv
\frac{\sigma_\parallel - \sigma_\perp}{\sigma_\parallel + \sigma_\perp} =
\frac{1}{p_\gamma} \frac
{W^L(0,\frac{\pi}{2}, \frac{\pi}{2})- W^L(0,\frac{\pi}{2}, 0)}
{W^L(0,\frac{\pi}{2}, \frac{\pi}{2})+ W^L(0,\frac{\pi}{2}, 0)},
\label{Sigma}
\end{equation}
where $\sigma_\parallel$ ($\sigma_\perp$) is the cross section for the
symmetric decay of particle pairs produced parallel (normal) to the plane
of polarization of the photon. 
$W^L$ represents angular distribution for the decay (\ref{decay}) with
linearly polarized photon beams.
In terms of the density matrix, it can be written as
\begin{equation}
\Sigma = \frac{\rho^1_{11} + \rho^1_{1-1}}{\rho^0_{11} +
\rho^0_{1-1}}.
\end{equation}

Another relevant quantity is the parity asymmetry $P_\sigma$,
which is defined from the observation that, if either natural
($P=\sigma$) or unnatural parity ($P=-\sigma$) exchange in the
$t$-channel contributes, one has an additional symmetry \cite{CSM68},
\begin{equation}
T_{-\lambda_V \lambda_N', -\lambda_\gamma \lambda_N} =
\pm (-1)^{\lambda_V - \lambda_\gamma}
T_{\lambda_V \lambda_N', \lambda_\gamma \lambda_N},
\label{Tsym}
\end{equation}
from which we get
\begin{equation}
\rho^{0(N/U)}_{\lambda \lambda'} = \frac12 \left[
\rho^0_{\lambda\lambda'} \mp (-1)^\lambda \rho^1_{-\lambda \lambda'}
\right].
\end{equation}
This allows us to define the parity asymmetry by means of $\sigma^N$
and $\sigma^U$, which are the contributions of natural and unnatural
parity exchanges to the cross section respectively as
\begin{equation}
P_\sigma \equiv \frac{\sigma^N - \sigma^U}{\sigma^N + \sigma^U}
= 2 \rho_{1-1}^1 - \rho^1_{00}.
\label{Psigma}
\end{equation}
Therefore when we have only the natural parity exchange we get
$P_\sigma = +1$, while we obtain $P_\sigma = -1$ when only the
unnatural parity exchange contributes.

We apply the above formalism to $\rho$ and $\phi$ meson
photoproduction with polarized photon beams.
We denote the four-momenta of the initial proton by $p$, that of the
final proton by $p'$, the photon beam four-momentum by $q$, and that
of the produced vector meson by $q_V^{}$.
The matrix element for the soft Pomeron exchange part reads 
\cite{DL84,LM95,PL97}
\begin{eqnarray}
T^P_{\lambda_V m',\lambda_\gamma m} &=&
i 12 \sqrt{4\pi\alpha_{\rm em}} \beta_u G_P (w^2,t)
F_1(t) \frac{m_V^2 \beta_f}{f_V}
\frac{1}{m_V^2 - t}
\left( \frac{2\mu_0^2}{2 \mu_0^2 + m_V^2 - t} \right)
\nonumber \\ && \mbox{} \times
\Bigl\{ \bar{u}_{m'} (p') \not \! q \  u_m(p)
\varepsilon_V^*(\lambda_V)
\cdot \varepsilon_\gamma (\lambda_\gamma)
- \left[ q \cdot \varepsilon_V^* (\lambda_V) \right]
\bar{u}_{m'} (p') \gamma_\mu u_m(p)
\varepsilon^\mu_\gamma (\lambda_\gamma)
\Bigr\},
\nonumber \\
\label{Tpom}
\end{eqnarray}
where the vector meson and the photon helicities are denoted by
$\lambda_V$ and $\lambda_\gamma$ while $m$ and $m'$ are the spin
projections of the initial and final nucleon, respectively.
The remaining quantities are defined by
\begin{eqnarray}
G_P (w^2,t) &=& \left( \frac{w^2}{s_0} \right)^{\alpha_P^{} (t) - 1}
\exp\left\{ - \frac{i\pi}{2} [ \alpha_P^{} (t) - 1 ] \right\},
\nonumber \\
F_1(t) &=& \frac{4m_p^2 - 2.8t}{(4m_p^2 - t)(1-t/0.71)^2},
\end{eqnarray}
with $w^2 = (2W^2 + 2 m_p^2 - m_V^2)/4$ and $W^2 = (p+q)^2$.
$m_p$ represents the proton mass, while $m_V$ the vector meson masses,
and the Pomeron trajectory is $\alpha_P^{}(t) = 1.08 + \alpha_P' t$
with $\alpha_P' = 1/s_0 = 0.25$ GeV$^{-2}$.
We use $\mu_0^2 = 1.1$ GeV$^2$, $\beta_u=\beta_d = 2.07$ GeV$^{-1}$
and $\beta_s = 1.45$ GeV$^{-1}$.
The vector meson decay constant is represented by $f_V$.

The $f_1$ exchange amplitude reads \cite{KMOV99}
\begin{eqnarray}
T^{f_1}_{\lambda_V m',\lambda_\gamma m} &=& i
g_{f_1 V \gamma}^{} g_{f_1NN}^{} F_{f_1NN}^{}(t) F_{f_1 V
\gamma}^{}(t)
\frac{m_V^2}{t-m_{f_1}^2}
\epsilon_{\mu\nu\alpha\beta} q^\mu \varepsilon_V^{*\nu}(\lambda_V)
\varepsilon^\alpha_\gamma (\lambda_\gamma)
\nonumber \\ && \mbox{} \times
\left( g^{\beta\delta} - \frac{(p-p')^\beta (p-p')^\delta}{m_{f_1}^2}
\right)
\bar{u}_{m'} (p') \gamma_\delta \gamma_5 u_{m} (p),
\label{Tf1}
\end{eqnarray}
where the $f_1 V \gamma$ coupling constants are determined from the
$f_1$ decay: $g_{f_1 \rho^0 \gamma} = 0.94$ GeV$^{-2}$ and
$g_{f_1 \phi \gamma} = 0.18$ GeV$^{-2}$.
The form factors are $F_{f_1NN}^{}(t) = 1/(1-t/m_{f_1}^2)^2$ with 
$m_{f_1}^{}$ ($=1.285$ GeV) defining the $f_1$ mass and 
$F_{f_1 V \gamma}^{}(t) = (\Lambda_V^2 - m_{f_1}^2)/(\Lambda_V^2-t)$
with $\Lambda_\rho = 1.5$ GeV and
$\Lambda_\phi = 1.8$ GeV.
We refer for the details of the amplitudes to Ref. \cite{KMOV99}.

\begin{figure}
\mbox{} \vskip 0.5cm
\centering
\epsfig{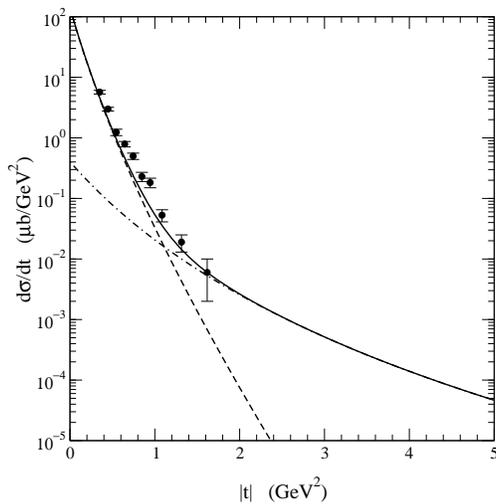}
\caption{The differential cross section for $\rho$ meson
photoproduction at $W = 94$ GeV.
The dashed and dot-dashed lines are the contributions from the Pomeron
and $f_1$ exchange, respectively, while the solid line is obtained by
including both exchanges.
Experimental data are from Ref. \protect\cite{ZEUS99}.}
\label{fig:rhot}
\end{figure}

In Fig. \ref{fig:rhot} we show the differential cross section for
$\rho$ photoproduction at $\gamma p$ c.m. energy $W = 94$ GeV, which
is the kinematical region of the ZEUS experiments.
The different role of the Pomeron and $f_1$ exchanges is apparent:
The Pomeron dominates at small $|t|$ while the $f_1$ 
gives the major contribution at larger $|t|$.
The differential cross section for $\phi$ photoproduction can be 
found in Ref. \cite{KMOV99}.

\begin{figure}
\mbox{} \vskip 1.0cm
\centering
\epsfig{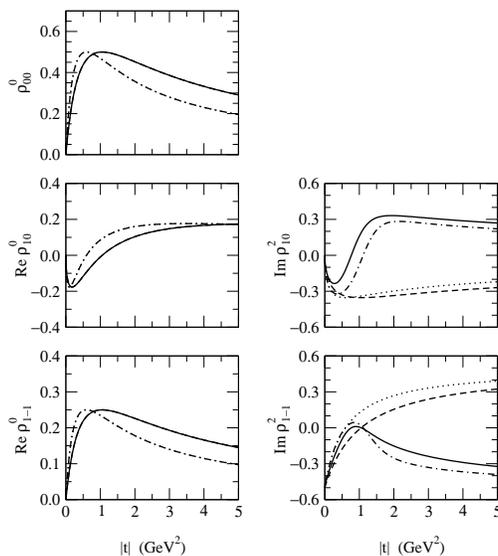}
\caption{Vector meson density matrix $\rho^{0,2}_{ik}$ in the
Gottfried-Jackson frame. The dotted and
dot-dashed lines (dashed and solid lines) are predictions of the
Pomeron exchange and Pomeron plus $f_1$ exchange models, respectively,
for $\rho$ ($\phi$) photoproduction at $W = 94$ GeV.
In the case of $\rho^0_{ik}$ (left panel) the two models give the
same results.}
\label{fig:rho02}
\end{figure}

\begin{figure}
\mbox{} \vskip 1.0cm
\centering
\epsfig{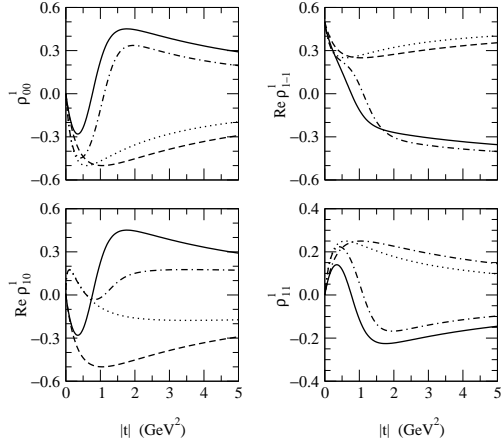}
\caption{Vector meson density matrix $\rho^{1}_{ik}$. Notations are
the same as in Fig. \ref{fig:rho02}.}
\label{fig:rho1}
\end{figure}

Figures \ref{fig:rho02} and \ref{fig:rho1} show the density matrices
defined in Eq. (\ref{dmdef}) for $\rho$ and $\phi$ photoproduction
for the same energy.
The figures emphasize the diverse features of the Pomeron and $f_1$
exchanges arising as a consequence from their different symmetry
properties (\ref{Tsym}).
The inclusion of the $f_1$ exchange changes the signs of some 
density matrix elements at large $|t|$ where the $f_1$ exchange 
dominates the process. This feature is responsible for the 
dramatic difference in the asymmetries between the two approaches.

\begin{figure}
\mbox{} \vskip 1.0cm
\centering
\epsfig{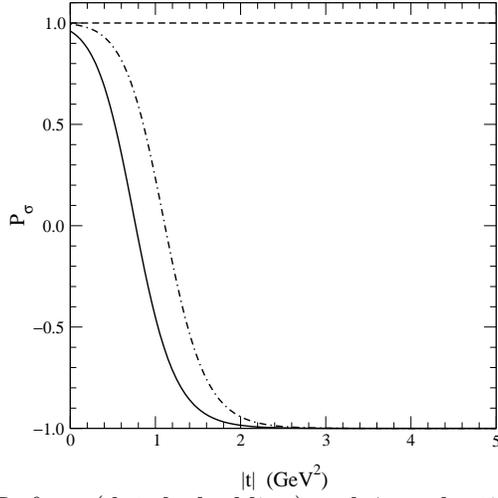}
\caption{The asymmetry $P_\sigma$ for $\rho$ (dot-dashed line) and
$\phi$ production (solid line) within Pomeron plus $f_1$ exchange.
The dashed line is the prediction of the Pomeron exchange for
$\rho$ and $\phi$ production.}
\label{fig:Psigma}
\end{figure}

We give predictions for the parity asymmetry $P_\sigma$ in
Fig.~\ref{fig:Psigma}.
We obtain identical result for the $\Sigma$ asymmetry of Eq. (\ref{Sigma}).
$\Sigma$ is not related unambiguously to natural and
unnatural parity exchange, but it becomes equivalent to $P_\sigma$
if the helicity-flip amplitudes are suppressed as in our case.
Furthermore, $P_\sigma = \pm 1$ implies $\Sigma = \pm 1$
although the reverse implication is not always true \cite{SSW70}.

Because of its natural parity, the Pomeron exchange leads to 
$P_\sigma = +1$ while the $f_1$ exchange gives $P_\sigma = -1$ 
due to its unnatural parity.
Therefore in Fig.~\ref{fig:Psigma} one can view the relative 
strength of the two exchanges as a function of $|t|$.
In $\rho$ photoproduction the two exchanges are comparable in
magnitude at $|t| \approx 1$ GeV$^2$, which leads to the vanishing of 
$P_\sigma$ in this region.
Below this region, the Pomeron dominates and the asymmetry approaches
$+1$, while it becomes $-1$ for $|t| > 2$ GeV$^2$ where the $f_1$
dominance is clearly established.

Although, as shown above, the best way to distinguish the two mechanisms
in vector meson production is to use the polarized photon beams, similar
information can be obtained from vector meson {\it electroproduction}
with unpolarized electron beam experiments \cite{SW73,WS78} at small $Q^2$,
which can be performed at present electron facilities.

Data on the density matrix in vector meson electroproduction by
fixed-target experiments is available \cite{JLMS76,SWGL82}.
Recently the H1 and ZEUS Collaborations \cite{H1-99,ZEUS99b} reported
data on the density matrix elements in $\rho^0$ electroproduction at
higher energy.
Both seem to be consistent with the Pomeron exchange model.
However it should be noted that these data were obtained only in the
region of small $|t|$, say $|t| \le 0.6$ GeV$^2$, with large errors.
In this region the natural parity exchange (Pomeron exchange) dominates
and the $f_1$ exchange contribution is small, so it is not possible to
draw any definite conclusion on the effect of the $f_1$ exchange from
these limited data set.
Since the $f_1$ exchange alters the predictions of the Pomeron exchange
at large $|t|$, it is necessary to measure the $|t|$-dependence of the
density matrices up to $|t| \approx 2$ GeV$^2$, and this may clarify the
nature of the exchanged trajectory which is responsible for vector meson
production at large $|t|$.

In summary, we have shown that the new anomalous unnatural-parity
$f_1$ exchange leads to significant $|t|$ dependence of the $P_\sigma$
and $\Sigma$ asymmetries in polarized vector meson photoproduction. 
The recent claim of Donnachie and Landshoff \cite{DL98,DL99} that
the relatively large $|t|$ data of the ZEUS experiments could be
explained by including the hard Pomeron will lead to a very different
prediction on these asymmetries and can be discriminated from the
$f_1$ exchange process.
We have good reason to believe that the existence of the anomalous
$f_1$ exchange in vector meson production is deeply related to the
properties of the axial anomaly in QCD \cite{KMOV99}.
Therefore the investigation of the decay asymmetries in vector meson
production by polarized photon or (un)polarized lepton beams at present
experimental facilities such as CERN, DESY and Fermilab will shed light
on our understanding of the diffractive processes from the fundamental
structure of QCD.

\bigskip

We are grateful to J.~A. Crittenden, S.~B. Gerasimov, and A.~I. Titov
for fruitful discussions.
Y.O. and D.P.M. were supported in part by
the KOSEF through the CTP of Seoul National University.
V.V. was supported by DGICYT-PB97-1227. 
N.I.K. thanks the Department of Physics of Seoul National 
University for the warm hospitality.

\newpage

\begin{references}


\bibitem{BSYP78}
T.~H. Bauer, R.~D. Spital, D.~R. Yennie, and F.~M. Pipkin,
  Rev. Mod. Phys. {\bf 50}, 261 (1978), (E) {\bf 51}, 407 (1979).

\bibitem{Crit97}
J.~A. Crittenden,
  {\it Exclusive production of neutral vector mesons at the
  electron-proton collider HERA\/},
  Vol. 140 of {\em Springer Tracts in Modern Physics\/},
  (Springer-Verlag, Berlin, 1997).

\bibitem{AC99}
H.~Abramowicz and A.~Caldwell,
  Rev. Mod. Phys. {\bf 71}. 1275 (1999).

\bibitem{DL87a}
A.~Donnachie and P.~V. Landshoff,
  Phys. Lett. B {\bf 185}, 403 (1987).

\bibitem{DL88a}
A.~Donnachie and P.~V. Landshoff,
  Phys. Lett. B {\bf 207}, 319 (1988).

\bibitem{Saku69}
J.~J. Sakurai,
  Phys. Rev. Lett. {\bf 22}, 981 (1969).

\bibitem{BCGG72}
J.~Ballam {\em et~al.\/},
  Phys. Rev. D {\bf 5}, 545 (1972).

\bibitem{BCEK73}
J.~Ballam {\em et~al.\/},
  Phys. Rev. D {\bf 7}, 3150 (1973).

\bibitem{ZEUS99}
\mbox{ZEUS Collaboration,} J.~Breitweg {\em et~al.\/},
  DESY Report No. DESY-99-160 (1999), hep-ex/9910038.

\bibitem{DL98}
A.~Donnachie and P.~V. Landshoff,
  Phys. Lett. B {\bf 437}, 408 (1998).

\bibitem{DL99}
A.~Donnachie and P.~V. Landshoff,
  Manchester Univ. Report No. MC-TH-99-16 (1999), hep-ph/9912312.

\bibitem{KMOV99}
N.~I. Kochelev, D.-P. Min, Y.~Oh, V.~Vento, and A.~V. Vinnikov,
  Seoul Nat'l Univ. Report No. SNUTP-99-048 (1999), hep-ph/9911480,
  to be published in Phys. Rev. D.

\bibitem{KL99}
D.~Kharzeev and E.~Levin,
  BNL Report No. BNL-NT-99-8 (1999), hep-ph/9912216.

\bibitem{Shu00}
E.~V. Shuryak,
  SUNY Stony Brook Report (2000), hep-ph/0001189.

\bibitem{GJ64}
K.~Gottfried and J.~D. Jackson,
  Nouvo Cimento {\bf 33}, 309 (1964).

\bibitem{SSW70}
K.~Schilling, P.~Seyboth, and G.~Wolf,
  Nucl. Phys. {\bf B15}, 397 (1970), (E) {\bf B18}, 332 (1970).

\bibitem{SW73}
K.~Schilling and G.~Wolf,
  Nucl. Phys. {\bf B61}, 381 (1973).

\bibitem{WS78}
G. Wolf and P. S{\"o}ding,
  in {\it Electromagnetic Interactions of Hadrons\/}, Vol. 2,
  edited by A.~Donnachie and G.~Shaw, (Plenum Press, New York, 1978) p. 1.

\bibitem{CSM68}
G.~Cohen-Tannoudji, \mbox{Ph}. Salin, and A.~Morel,
  Nouvo Cimento {\bf 55}, 412 (1968).

\bibitem{DL84}
A.~Donnachie and P.~V. Landshoff,
  Nucl. Phys. {\bf B244}, 322 (1984).

\bibitem{LM95}
J.-M. Laget and R.~Mendez-Galain,
  Nucl. Phys. {\bf A581}, 397 (1995).

\bibitem{PL97}
M.~A. Pichowsky and T.-S.~H. Lee,
  Phys. Rev. D {\bf 56}, 1644 (1997).

\bibitem{JLMS76}
P.~Joos {\em et~al.\/}, 
  Nucl. Phys. {\bf B113}, 53 (1976).

\bibitem{SWGL82}
W.~D. Shambroom {\em et~al.\/},
  Phys. Rev. D {\bf 26}, 1 (1982).

\bibitem{H1-99}
\mbox{H1 Collaboration,} C.~Adloff {\em et~al.\/},
  DESY Report No. DESY-99-010 (1999), hep-ex/9902019.

\bibitem{ZEUS99b}
\mbox{ZEUS Collaboration,} J.~Breitweg {\em et~al.\/},
  DESY Report No. DESY-99-102 (1999), hep-ex/9908026.


\end{references}
\end{document}